# Effects of different tumors on the steady-state heat distribution in the human eye using the 3D finite element method

Aboozar Moradi

Department of Electrical Engineering,
University of Tehran
Tehran, Iran

a.moradi@ut.ac.ir

Mohammad Reza Yousefi

Digital processing and machine vision research center,
Najafabad Branch, Azad University
Isfahan, Iran

mr-yousefi@iaun.ac.ir

**Abstract-** In this paper, a three-dimensional finite element method is developed to simulate the heat distribution in the human eye with different types of tumors to understand the effect of tumors on heat distribution in the human eye. The human eye is modeled as a composition of several homogeneous regions and the physical and thermal properties of each region used in this study are more accurate than the models used in previous studies. By considering the exact and complicated geometry of all parts, the finite element method is a proper solution for solving the heat equation inside the human eye. There are two kinds of boundary conditions called the radiation condition and the Robin condition. The radiation boundary condition is modeled as a Robin boundary condition. For modeling eye tumors and their effect on heat distribution, we need information about eye tumor properties such as heat conductivity, density, specific heat, and so on. Thanks to no accurate reported information about eye tumor properties, the properties of other types of tumors such as skin, and bowel tumors are used. Simulation results with different parameters of eye tumors show the effect of eye tumors on heat distribution in the human eye.

## 1. Introduction

The tumor may cause a considerable effect on heat distribution inside the human eye, and it increases the temperature in its surrounding tissues, and because of the sensitivity of eye tissues and cells to heat, increasing the temperature inside the eye may damage the eye tissues and cells. Therefore, by modeling tumors in the human eye, we can predict and understand the heat distribution inside the human eye. Various mathematical models that describe heat distribution using different methods are reported in the literature.

In Scot's model [1], [2], the human eye was modeled as the composition of 7 homogenous regions and a two-dimensional model of heat distribution in the human eye was solved using the finite element method. A three-dimensional boundary element model of the human eye was developed to investigate the thermal effects of eye tumors on the ocular temperature distribution in [3]. In [4], a boundary element method was applied for the numerical solution of a boundary value problem for a two-dimensional steady-state bioheat transfer model of the human eye. In [5], a simple finite element 2D model of the human eye was developed to calculate the steady-state temperature distribution. In [6], the finite element method was used for the numerical simulation of the two-dimensional transient bioheat transfer process in the human eye. In [7]–[10], the finite element method was used for the numerical simulation of the three-dimensional transient transfer process in the human eye with and without tumors. In this paper, we investigate the effect of tumors on heat distribution in the human eye.

Despite previous studies, in this paper, we use three-dimensional finite element method to simulate heat distribution in the human eye with tumors. Therefore, we use the proposed method in [1] to model the radiation boundary condition on the cornea surface to the Robin boundary condition. The physical properties of tumors in this paper are like the properties used are previous studies [3], [7]–[9].

## 2. Description of the human eye tissue

A good representation of the human eye and associated tissues has been taken from Gray's Anatomy. In Fig 1(a), we can see that the eye is approximately a spherical organ. The protective outer layer of the eye, sometimes referred to as "the white of the eye" is called the sclera and it maintains the shape of the eye. The front portion of the sclera is called the cornea, and this is transparent and allows light to enter the eye. The cornea is a powerful refracting surface that provides much of the eye's focusing power. The choroid is the second layer of the eye and lies between the sclera and the retina. It contains the blood vessels that provide nourishment to the outer layers of the retina. The iris is part of the eye that gives it color. It consists of muscular tissue that responds to surrounding light, making the pupil or circular opening in the center of the iris, larger or smaller depending on the brightness of the light. Light entering the pupil falls onto the lens of the eye where it is altered before passing through to the retina. The lens is a transparent, biconvex structure, encased in a thin transparent covering. The function of the lens is to refract and focus incoming light onto the retina for processing. The retina is the innermost layer of the eye. It converts images into electrical impulses that are sent along the optic nerve to the brain where the images are interpreted [11]–[16]. Inside of the eyeball is divided by the lens into two fluid-filled sections. The larger section at the back of the eye is filled with a colorless gelatinous mass called vitreous humor. The smaller section in the front contains a clear, water-like material called aqueous humor. Melanoma is a kind of malignant tumor that the probability of its existence in the skin, bowel, and eye is more than in other regions of the human body [17]–[19]. In regions such as sunny climates, the probability of the existence of melanoma is high and for Negro and Asian people the probability of existence is low. In this paper, we will focus only on choroidal melanoma. The melanomas are classified based on the tumor thickness and penetration depth in the tissue and they will express in millimeters.

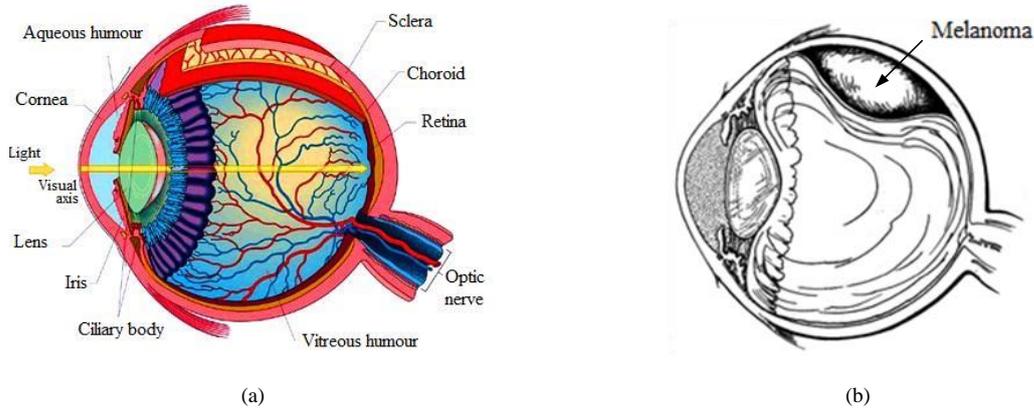

(a) (b)
Fig 1. Human eye (a) normal eye [20] (b) human eye with tumor [17]

There are three kinds of tumors based on size called $T_1$, $T_2$, and $T_3$. As shown in Fig 1(b), the location of these tumors for simulation is between the sclera and vitreous homour, and their properties are shown in table 1. In tumor $T_1$, the thickness of the tumor is between 1-3 millimeters and the length is less than 10 millimeters, for tumor $T_2$, the thickness is between 3-5 millimeters and the length is between 10-15 millimeters and for $T_3$, the thickness is more than 5 millimeter, and the length is more than 15 millimeters [17].

Table 1. choroidal melanoma classification based on T category [17].

| Category | Largest basal diameter (mm) | Thickness (mm) |
|---|---|---|
| $T_1$ | $\leq 10$ | $\leq 3$ |
| $T_2$ | $10 - 15$ | $3 - 5$ |
| $T_3$ | $> 15$ | $> 5$ |

## 3. Mathematical model

In this paper, a simple 3D model of the human eye with a tumor is developed to investigate heat transfer process in the human eye. Some simplifications are adopted regarding geometry and modeling process such as the eye is a solid structure consisting of various tissues in contact with each other. Therefore, the eye is divided into eight regions that each region is assumed homogeneous. Several heat transfer equations for living tissues have been developed. In this paper, we use the Pennes heat equation as follows [6, 7]:

$$\partial/\partial x(k_{i_x}\partial T_i/\partial x)+ \partial/\partial y(k_{i_y}\partial T_i/\partial y)+ \partial/\partial z(k_{i_z}\partial T_i/\partial z)+ Q_i = \beta T_i \quad i=1, 2... 7 \quad (1)$$

Where k is the thermal conductivity, $T$ is the temperature, $Q$ is generalized source term and $\beta$ is a constant depends on the source of heat transfers between different regions (tissues). The index $i$ indicates a region of the human eye, in particular the cornea, aqueous humour, iris, lens, vitreous, sclera, and optic nerve. In a normal eye (without tumor), no heat source is considered and $Q = 0$ and it is assumed $\beta = 0$. The existence of tumors in the human eye causes that Q and will have nonzero values that will be shown in section 5 and table 3. The physical parameters for regions are in table 2. Therefore, the heat equation can be expressed:

$$\partial/\partial x(k_{i_x} \partial T_i/\partial x) + \partial/\partial y(k_{i_y} \partial T_i/\partial y) + \partial/\partial z(k_{i_z} \partial T_i/\partial z) = 0 \qquad (2)$$

Table 2. Thermal physical parameters of a human eye

| Region | $C[J/(KgK)]$ | $\rho[Kg/m^3]$ | $k[W/(mK)]$ |
|---|---|---|---|
| Cornea | 4178[12] | 1050[11] | 0.58[10] |
| Iris | 3997[6,7] | 1000[6,7] | 1.0042[6,7] |
| Aqueous humour | 3997[12] | 1000[12] | 0.58[10] |
| Lens | 3000[12] | 1050[11] | 0.40[11] |
| Sclera | 4178[12] | 1000[12] | 0.603[12] |
| Vitreous | 4178[12] | 1000[12] | 0.603[12] |
| Optic nerve | 3997[6,7] | 1000[6,7] | 1.0042[6,7] |
| Thermal conductivity of tumor (Wm$^{-1}$ K$^{-1}$) | 0.35–0.67 [2] | | |
| Tumor metabolic heat generation of tumor (Wm−3) | 15,000-80,000 [2] | | |
| Tumor blood perfusion of tumor rate (m$^3$ s$^{-1}$ m$^{-3}$) | 0.0014–0.0072 [2] | | |

## 4. Boundary conditions

The boundary conditions used in this study are like the ones used for the 2D model [6]. Two kinds of boundary conditions are used. The first boundary condition is applied on the corneal surface named that is shown in figure 2, where three heat loss mechanisms are due to convection, radiation and heat loss as a result of tear evaporation from the corneal surface takes place. Mathematically, this may be written as [7]–[9]:

$$k_1 \partial T_1 / \partial n = \alpha_a (T_1 - T_a) + \varepsilon\sigma(T_1^4 - T_a^4) + E; \quad on \; S_1 \qquad (3)$$

Where $k_1$ is the thermal conductivity of cornea, $\alpha_a = 10W/(m^2 K)$ is heat transfer coefficient between cornea and environment, $T_a = 25^oC$ is the temperature of surrounding environment, $\varepsilon = 0.975W/m^3$ is the corneal emissivity, $\sigma = 5.67 \times 10^{-8} W/(m^2 K^4)$ is the Stefan-Boltzmann constant and $E = 40W/m^2$ is the loss of heat flux due to the evaporation of tears.

The second boundary condition is applied on the scleroid surface and named that is shown in figure 2. The human eye is embedded in homogeneous surrounding anatomy, which is at body core temperature. Consequently, heat transfer from the surrounding environment to the eye may be described by a single heat transfer coefficient as follows:

$$k_2 \partial T_2 / \partial n = \alpha_{bl}(T_2 - T_{bl}); \quad on \ S_2 \tag{4}$$

Where the thermal conductivity of the sclera is, $\alpha_{bl} = 65W/(m^2 K)$ is the heat transfer coefficient between sclera and blood vessels and $T_b = 37^o C$ is the temperature of the blood.

The boundary condition is shown in Fig.2, the red region refers to the corneal surface and the blue region refers to the scleroid surface.

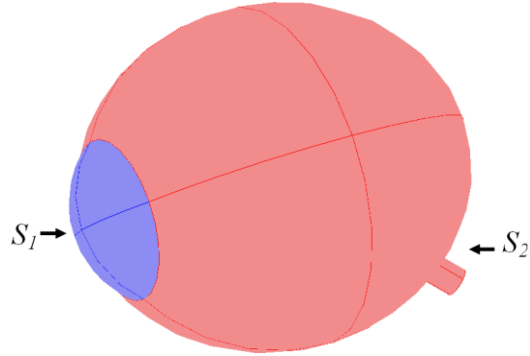

Fig 2. Two boundary conditions on the human eye surface

## 5. Method

This paper is based on the Ritz method in the finite element method domain, in the finite element method we divide the region of interest into some elements. Therefore, we try to solve the heat equation in each element and finally we solve the heat equation in our model. In Fig. 3, we can see seven regions and a tumor for 3-dimensional and the result of dividing our 3-dimensional model into small elements (apply to mesh). The Ritz method, also known as the Rayleigh-Ritz method, is a variational method in which the boundary-value problem is formulated in terms of a variational expression, called functional. The minimum of this function corresponds to the governing differential equation

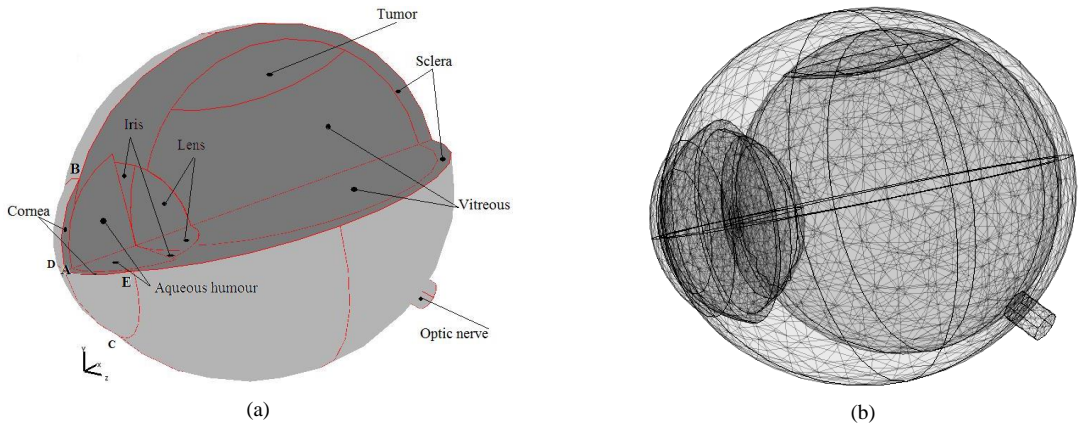

Fig 3. Human eye model (a) three dimensional (b) apply meshing in 3D model of the human eye. The eye model consists of 11720 triangular elements, 2195 nodes and 7 subdomains.

Under the given boundary conditions, the approximate solution is then obtained by minimizing the functional variables that define a certain approximation to the solution. For seven regions in table 2, there are not any blood vessels, therefore, the steady state general heat equation for these regions is expressed in the following equation:

$$k\nabla^2 T + Q = 0 \qquad (5)$$

When we have a tumor in the human eye, the heat equation for the tumor region is expressed based on Penne's equation in the following equation:

$$k\nabla^2 T + Q = P \qquad (6)$$

Where Q is the heat source. In a normal eye Q is zero. In the above equation, P is in Penne's equation, so $\omega_b$, $c_b = 1.02 \times$ specific heat of water $= 4270$ and $T_b = 310.15$ Kelvin are blood flow rate, specific heat, and blood temperature [16, 17]. Finally, we can develop the three-dimensional heat equation for all regions in the human eye in the following form:

$$\partial/\partial x\left(k_{i_x}\partial T_i/\partial x\right) + \partial/\partial y\left(k_{i_y}\partial T_i/\partial y\right) + \partial/\partial z\left(k_{i_z}\partial T_i/\partial z\right) + Q_i = \beta_i T_i \qquad (7)$$

where i=1…8 including seven homogenous regions and tumor. In the above equation, for the tumor region we have:

$$Q_i = \omega_b c_b T_b \qquad (8)$$

$$\beta_i = \omega_b c_b$$

We use the Ritz method to form a functional equation as:

$$F(T) = 1/2 \iiint_v \left[k\left((\partial T/\partial x)^2 + (\partial T/\partial y)^2 + (\partial T/\partial z)^2\right) + (\beta T)T\right]dV - \iiint_v QTdV \qquad (9)$$

The above equation is before applying boundary conditions. Now we must solve (9) for each element and finally after minimizing the functional for each element, we obtain a global form matrix for our model (all elements) in the following form:

$$KT = B \qquad (10)$$

A complete explanation is presented in appendix A.

## 6. Incorporation of boundary condition

As we know, we have two different boundary conditions. To apply boundary conditions in equations (3), and (4), we use the following method:

**Step 1:** Compose the functional for equation (3)

$$F_b(T) = \iint_{s_1} (\alpha_{bl}/2 \times T^2 - (\alpha_{bl} T_{bl})T) ds \qquad (11)$$

**Step 2:** Compose the matrix form of equation (9) and apply it to the system matrices

$$T^s = \sum_{j=1}^{3} N_j^s T_j^s \tag{12}$$

we have:

$$\partial F_b^s / \partial T_i^s = \sum_{j=1}^{3} T_j^s (\iint_{s_1} \alpha_{bl} N_i^s N_j^s ds - \iint_{s_1} \alpha_{bl} T_{bl} N_i^s ds) \tag{13}$$

Equation (13) can be written as:

$$\{\partial F_b^s / \partial T_i^s\} = [K^s]\{T^s\} - \{b^s\} \tag{14}$$

Therefore:

$$K_{ij}^s = \alpha_{bl} \Delta^s / 12 \times (1 + \delta_{ij}) \tag{15}$$

$$b_i^s = \alpha_{bl} \Delta^s / 3 \times T_{bl} \tag{16}$$

**Step 3:** Assembling [$K^s$] for each segment into a global form of [K] and global form of {b} can be accomplished easily by adding each $K_{ij}^s$ to $K_{ns(i,s),ns(j,s)}$ and $b_i^s$ to $b_{ns(i,s)}$.

**Step 4:** We compose the functional for equation (4)

$$F_b(T) = \iint_{s_2} (\alpha_a / 2 \times T^2 + \varepsilon\sigma / 2 \times T^5 - (E + \alpha_a T_{liquid} + \varepsilon\sigma(T_\infty)^4)T)ds \tag{17}$$

To minimize the above functional, we calculate the derivative of the above equation to T, this leads to a nonlinear equation and complicated matrix form for solving. For simplification, we used the method proposed in [1] to model the radiation boundary condition to a robin boundary condition with coefficient $h_{rad}$ as shown in the following equations

$$k_1 \partial T_1 / \partial n = \alpha_a (T_1 - T_a) + \varepsilon\sigma(T_1 + T_a)(T_1^2 + T_a^2)(T_1 - T_a) + E \tag{18}$$

$$k_1 \partial T_1 / \partial n = \alpha_a (T_1 - T_a) + h(T_1 - T_a) + E \tag{19}$$

$$h = \varepsilon\sigma(T_1 + T_a)(T_1^2 + T_a^2) \tag{20}$$

$$k_1 \partial T_1 / \partial n = h_{rad}(T_1 - T_a) + E \tag{21}$$

$$h_{rad} = \alpha_a + h \tag{22}$$

In the above equation h=6.

## 7. Results

For the implementation of the 3D model of the eye, in the first step, we draw all parts of the model, so we apply boundary conditions, mesh the model, and finally, we solve Eqn. (10). Steady-state temperature variation along the visual axis for 3D used in this paper and the 2D model in [2] is shown in Fig 4. Comparing the results of 2D and 3D models in the steady state shows that the results are close to each other. Since 2D simulation assumed the heat distribution in one direction is negligible, the results of 3D models are more accurate than the 2D models. Since blood temperature is $37^0C$, in the steady state, the temperature of all parts of the eye except the cornea which is in contact with environment must be close to $37^0C$. Fig 4 shows that the temperature of all parts of human eye for the 3D and 2D models are about 36.9950 $^0C$ and 36.7840 $^0C$, respectively. Therefore, the results of the 3D model are more accurate [6].

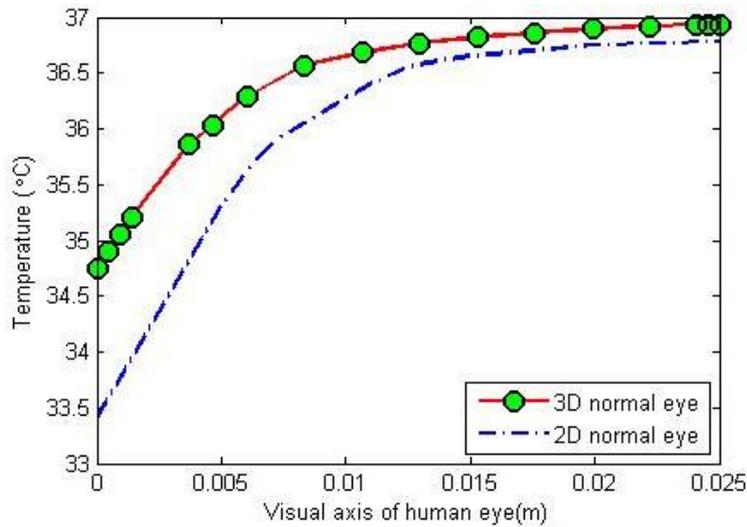

Fig 4. Temperature variation along visual axis for 3D used in this paper and 2D model in for steady state.

For tumors in human eye, there is not any accurate report about the properties of eye tumors. Therefore, to obtain complete results for heat distribution inside the human eye, we choose different values for thermal conductivity, blood flow rate, and metabolic heat generation which are used for other tumors inside the body [17].

From the measurements compiled in [13], the thermal conductivity of tumors in various parts of the human body was found to be in the range of 0.35 to 0.67 $Wm^{-1}K^{-1}$.

The types of tumors that were measured included tumors of the breast, colon, liver, lungs, and pancreas, and the center of the elliptical used in generating the tumor is placed at coordinates (0.015, 0.0075, 0). The values of the blood flow rate of tumors that were measured by various researchers using different techniques were compiled in [14]. Based on the data measured on breast tumors, lymphomas, anaplastic, carcinoma, and differentiated tumors, the blood flow rate of the tumor was found between 0.0014 and 0.0072 $m^3s^{-1}m^{-3}$. Therefore, because there is not any information on tumor in the human eye, we used the properties of other kinds of tumors in the human body.

To investigate the temperature changes inside the human eye, in table 2 & 3 we can see different parameters that are used for the simulation heat transfer in the human eye. Results are given in Fig 5 for the eye with tumor T1. For simulation, we must choose different values for the coefficient in equation (7).

In table 3, we can see the results of different tumor metabolic heat generation, blood flow rate, and Thermal conductivity on $Q_i$ and $\beta_i$ in equation (8). Fig. 5 is for differentiating between the temperature on the visual axes of the normal and abnormal eye with tumor T1. In each diagram, one of the three parameters in table 3 varies and two other parameters are constant. By comparison these diagrams, we can understand that the maximum difference between the temperature on the visual axes of the normal and abnormal eye is for the situations in which metabolic heat generation changes and it equals 75000.

In Fig. 6, we can compare the difference between temperature in the normal and abnormal eye (in tumors T1, T2, and T3) for blood flow rate=0.006 ($m^3$ $s^{-1}$ $m^{-3}$), thermal conductivity=0.5 ($Wm^{-1}$ $K^{-1}$), and metabolic heat generation=45000 ($Wm^{-3}$) on the visual axis.

Table 3. Different values of various value of tumor, Thermal conductivity, Blood perfusion rate and metabolic heat generation of tumor and their effect on $Q_i$ and $\beta_i$ in equation (7).

| Various value of tumor | Tumor metabolic heat generation $Q_m$ ($Wm^{-3}$) | Tumor blood perfusion rate $\omega_b$ ($m^3$ $s^{-1}$ $m^{-3}$) | Thermal conductivity $k(Wm^{-1}$ $K^{-1})$ | coefficient in equation (8) | |
| --- | --- | --- | --- | --- | --- |
| | | | | $Q_i$ | $\beta_i$ |
| Thermal conductivity | 45000 | 0.006 | 0.3 | 8467805.58 | 27157.2 |
| | 45000 | 0.006 | 0.4 | 8467805.58 | 27157.2 |
| | 45000 | 0.006 | 0.5 | 8467805.58 | 27157.2 |
| | 45000 | 0.006 | 0.6 | 8467805.58 | 27157.2 |
| | 45000 | 0.006 | 0.7 | 8467805.58 | 27157.2 |
| Blood perfusion rate | 45000 | 0.002 | 0.5 | 2852601.86 | 9052.4 |
| | 45000 | 0.004 | 0.5 | 5660203.72 | 18104.8 |
| | 45000 | 0.006 | 0.5 | 8467805.58 | 27157.2 |
| | 45000 | 0.008 | 0.5 | 11275407.44 | 36209.6 |
| | 45000 | 0.010 | 0.5 | 14083009.3 | 45262 |
| Metabolic heat generation | 15000 | 0.006 | 0.5 | 8437805.58 | 27157.2 |
| | 30000 | 0.006 | 0.5 | 8452805.58 | 27157.2 |
| | 45000 | 0.006 | 0.5 | 8467805.58 | 27157.2 |
| | 60000 | 0.006 | 0.5 | 8482805.58 | 27157.2 |
| | 75000 | 0.006 | 0.5 | 8497805.58 | 27157.2 |

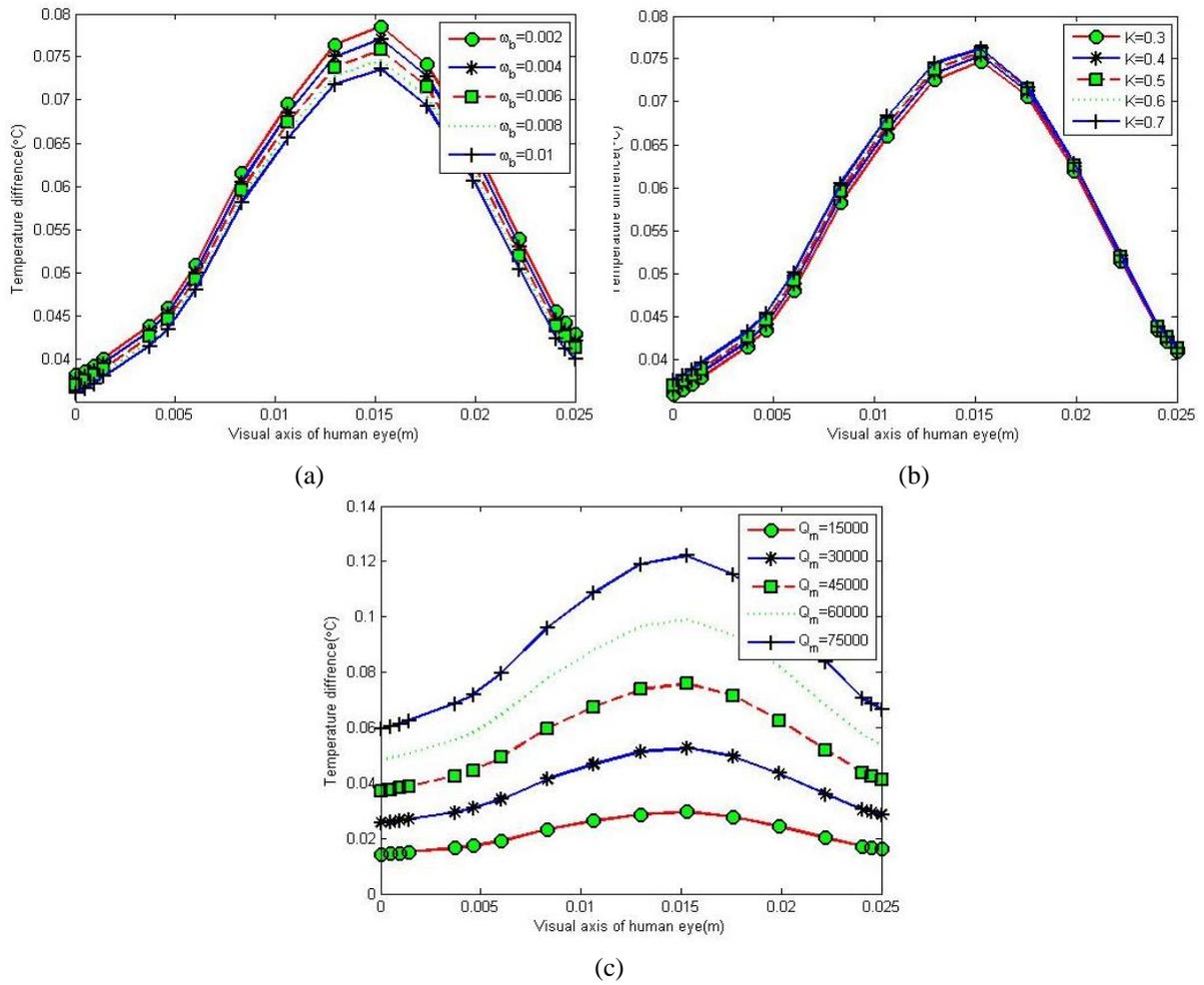

(c)

Fig 5. Temperature between normal and abnormal eye for category T1 tumor on paraxial for different values of (a) blood perfusion rate (b) thermal conductivity (c) metabolic heat generation

Therefore, if the tumor is larger, its effect on increasing the temperature of the human eye will be more.

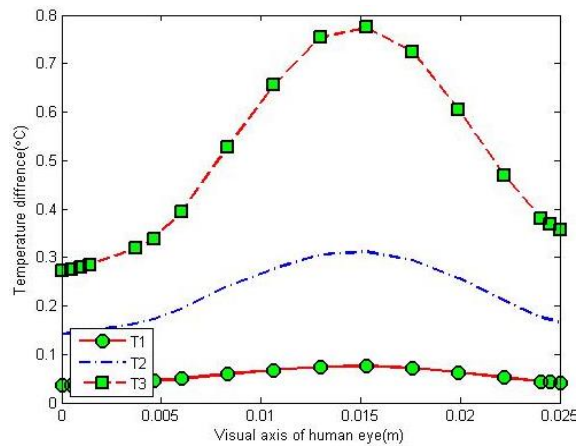

Fig 6. Comparison the temperature between three kinds of tumors between normal and abnormal eye.

## 7.1 Temperature on the corneal surface

To investigate the temperature changes on the corneal surface, five points, namely the central, superior, inferior, nasal, and temporal which we denote as A, B, C, D, and E, respectively, are selected on the corneal surface as illustrated in Fig. 3. Coordinates of each of these points are given by (0,0,0), (0.0025,0.00536,0), (0.0025, −0.00536, 0), (0.0025,0, −0.00536) and (0.0025,0,0.00536) respectively. Increase in temperature of these points for parameters that cause maximum effect in temperature of these points and central points of tumor are shown in table 4. In this state, results are obtained for values of eye tumor parameters like thermal conductivity, blood flow rate, and metabolic heat generation of $0.50 Wm^{-1}K^{-1}$, $0.006\ m^3 s^{-1} m^{-3}$ and $75,000\ Wm^{-3}$, respectively. For better comparison, the results for other values of eye tumor parameters like thermal conductivity, blood flow rate and metabolic heat generation with values of $0.50\ Wm^{-1}K^{-1}$, $0.006\ m^3 s^{-1} m^{-3}$ and $45,000\ Wm^{-3}$ are shown in table 5.

Table 4: Increase in temperature on the corneal surface and tumor center for an extreme case, in this state Thermal conductivity=0.5, Blood perfusion rate=0.006 and Metabolic heat generation=75000 for tumor

| Location | Normal | Temperature in Tumor category(°C) | | |
|---|---|---|---|---|
| | | $T_1$ | $T_2$ | $T_3$ |
| Central (A) | 34.742 | 34.8016 | 34.9688 | 35.1758 |
| Superior (B) | 35.9349 | 36.0153 | 36.2274 | 36.4399 |
| Inferior (C) | 35.9344 | 35.9806 | 36.1131 | 36.3013 |
| Nasal (D) | 35.9321 | 35.9906 | 36.1552 | 36.3634 |
| Temporal (E) | 35.9378 | 35.9958 | 36.1608 | 36.3682 |
| Center of tumor | ---- | 37.0556 | 37.3409 | 37.5626 |

Table 5: temperature on corneal surface and the tumor center and thermal conductivity=0.5, Blood perfusion rate=0.006 and Metabolic heat generation=45000 for tumor

| Location | Normal | Temperature in Tumor category(°C) | | |
|---|---|---|---|---|
| | | $T_1$ | $T_2$ | $T_3$ |
| Central (A) | 34.742 | 34.7789 | 34.8837 | 35.0143 |
| Superior (B) | 35.9349 | 35.9848 | 36.1178 | 36.252 |
| Inferior (C) | 35.9344 | 35.9631 | 36.0461 | 36.1646 |
| Nasal (D) | 35.9321 | 35.9684 | 36.0715 | 36.2027 |
| Temporal (E) | 35.9378 | 35.9736 | 36.0772 | 36.2077 |
| Center of tumor | ---- | 36.9877 | 37.1622 | 37.2971 |

## 8. Conclusion

In this paper, we presented the simulation results of heat distribution in the human eye with different types of tumor parameters, and we introduced a new method for modeling the radiation boundary condition for an accurate Robin boundary condition. Using the finite element method, we can solve heat equations in complicated eye geometry. There is no report for tumor parameters in humans. Therefore, we use tumor parameters that exist in the human body such as a tumor in the breast, colon, liver, lungs, and pancreas. The eye tumor geometry used in this paper is more accurate than the tumor geometry model used in [3]. The current numerical study has not been validated by any experiment works, but we can predict the presence of tumors by considering the obtained results for the normal and abnormal eyes. By considering the obtained results in the last figures, we can understand that tumors cause temperature increases because the tumor acts as a high heat source. Using the results in table 4 and 5, we can predict the presence of each kind of tumor that existed in the human eye. Most tissues and cells in the human eye are sensitive to heat temperature. Therefore, using these results we can predict heat distribution before it dramatically affects cellular mechanisms.

## A. Finite element method

For solving the heat equation, we use a linear tetrahedral element and the unknown temperature can be approximated as:

$$T^e(x, y, z) = a^e + b^e x + c^e y + d^e z \tag{A.1}$$

The coefficient $a^e$, $b^e$, $c^e$ and $d^e$ can be enforcing (A.1) at the four nodes of the element assuming a given value at the vertices. Thus, denoting the value of T at the jth node as $T_j^e$ we can express:

$$T_j^e(x, y, z) = a^e + b^e x_j^e + c^e y_j^e + d^e z_j^e \qquad j=1,2,3,4 \tag{A.2}$$

And we have:

$$a^e = \frac{1}{6v^e} \begin{bmatrix} T_1^e & T_2^e & T_3^e & T_4^e \\ x_1^e & x_2^e & x_3^e & x_4^e \\ y_1^e & y_2^e & y_3^e & y_4^e \\ z_1^e & z_2^e & z_3^e & z_4^e \end{bmatrix}$$

$$b^e = \frac{1}{6v^e} \begin{bmatrix} 1 & 1 & 1 & 1 \\ T_1^e & T_2^e & T_3^e & T_4^e \\ y_1^e & y_2^e & y_3^e & y_4^e \\ z_1^e & z_2^e & z_3^e & z_4^e \end{bmatrix}$$

$$c^e = \frac{1}{6v^e} \begin{bmatrix} 1 & 1 & 1 & 1 \\ x_1^e & x_2^e & x_3^e & x_4^e \\ T_1^e & T_2^e & T_3^e & T_4^e \\ z_1^e & z_2^e & z_3^e & z_4^e \end{bmatrix} \tag{A.3}$$

$$d^e = \frac{1}{6v^e}\begin{bmatrix} 1 & 1 & 1 & 1 \\ x_1^e & x_2^e & x_3^e & x_4^e \\ y_1^e & y_2^e & y_3^e & y_4^e \\ T_1^e & T_2^e & T_3^e & T_4^e \end{bmatrix}$$

Where:

$$v^e = \frac{1}{6v^e}\begin{bmatrix} 1 & 1 & 1 & 1 \\ x_1^e & x_2^e & x_3^e & x_4^e \\ y_1^e & y_2^e & y_3^e & y_4^e \\ z_1^e & z_2^e & z_3^e & z_4^e \end{bmatrix} \tag{A.4}$$

Thus we have

$$T^e(x,y,z) = \sum_{j=1}^{4} N_j^e(x,y,z) T_j^e \tag{A.5}$$

Where the interpolation function is given by

$$N_j^e(x,y,z) = 1/6v^e \times (a_j^e + b_j^e x + c_j^e y + d_j^e z) \tag{A.6}$$

It can be shown that the interpolation functions have the following property

$$N_j^e(x,y,z) = \delta_{ij} = \begin{cases} 1 & i=j \\ 0 & i \neq j \end{cases} \tag{A.7}$$

The functional for this case contains only the volume integral and it can be written as

$$F(T) = \sum_{e=1}^{M} F^e(T^e) \tag{A.8}$$

Where M denotes the total number of volume elements and $F^e$ is given by

$$F^e(T^e) = 1/2 \iiint_{v^e} \left[ k\left((\partial T^e/\partial x)^2 + (\partial T^e/\partial y)^2 + (\partial T^e/\partial z)^2\right) + (\beta^e T^e)T^e \right] dV - \iiint_{v^e} Q^e T^e dV \tag{A.9}$$

In (A.9) we didn't consider boundary conditions to calculate system matrices. In which $v^e$ denotes the volume of the eth element. Substituting (A.5) into (A.9) and taking the partial derivative of $F^e$ with respect to $T_j^e$, we obtain

$$\partial F^e/\partial T_i^e = \sum_{j=1}^{4} T_j^e \iiint_{v^e} \left[ k^e\left((\partial N_i^e/\partial x)(\partial N_j^e/\partial x) + (\partial N_i^e/\partial y)(\partial N_j^e/\partial y) + (\partial N_i^e/\partial z)(\partial N_j^e/\partial z)\right) + \beta^e N_i^e N_j^e \right] dV - \sum_{j=1}^{4} \iiint_{v^e} Q^e T^e dV \tag{A.10}$$

i=1,2,3,4

Now above equation can be written matrix form as

$$\{\partial F^e/\partial T^e\} = [K^e]\{T^e\} - \{b^e\} \tag{A.11}$$

Here

$$K_{ij}^e = \iiint_{v^e} \left[ k^e \left( (\partial N_i^e / \partial x)(\partial N_j^e / \partial x) + (\partial N_i^e / \partial y)(\partial N_j^e / \partial y) + (\partial N_i^e / \partial z)(\partial N_j^e / \partial z) \right) + \beta^e N_i^e N_j^e \right] dV$$

$$b_i^e = \iiint_{v^e} Q N_i^e dV \tag{A.12}$$

A basic formula to be used in this process is

$$\iiint_{v^e} (N_1^e)^k (N_2^e)^l (N_3^e)^m (N_4^e)^n dV = \frac{k!l!m!n!}{(k+l+m+n+3)!} 6V^e \tag{A.13}$$

And the result is

$$K_{ij}^e = \frac{k^e}{36 v^e} (b_i^e b_j^e + c_i^e c_j^e + d_i^e d_j^e) + \frac{v^e}{20} \beta^e (1 + \delta_{ij}) \tag{A.14}$$

$$b_i^e = \frac{V^e}{4} f^e$$